\documentclass[amssymb,prb,twocolumn,floats,amsmath,showpacs,superscriptaddress]{revtex4}
\usepackage[T1]{fontenc}
\usepackage{bm}
\usepackage{graphicx}
\usepackage{amssymb}
\usepackage{amsfonts}
\usepackage{amsmath}
\usepackage{multirow}

\begin{document}

\title{Relation between exciton splittings, magnetic circular dichroism, and magnetization in wurtzite (Ga,Fe)N}

\author{J.-G. Rousset}
\affiliation{Institute of Experimental Physics, Faculty of Physics, University of Warsaw, ul. Ho\.za 69, PL-00-681 Warszawa, Poland}
\author{J. Papierska}
\affiliation{Institute of Experimental Physics, Faculty of Physics, University of Warsaw, ul. Ho\.za 69, PL-00-681 Warszawa, Poland}
\author{W. Pacuski}
\affiliation{Institute of Experimental Physics, Faculty of Physics, University of Warsaw, ul. Ho\.za 69, PL-00-681 Warszawa, Poland}
\author{A. Golnik}
\affiliation{Institute of Experimental Physics, Faculty of Physics, University of Warsaw, ul. Ho\.za 69, PL-00-681 Warszawa, Poland}
\author{M. Nawrocki}
\affiliation{Institute of Experimental Physics, Faculty of Physics, University of Warsaw, ul. Ho\.za 69, PL-00-681 Warszawa, Poland}
\author{W. Stefanowicz}
\affiliation{Institute of Physics, Polish Academy of Sciences, al. Lotnik\'{o}w 32/46, 02-668 Warszawa, Poland}
\author{S. Stefanowicz}
\affiliation{Institute of Physics, Polish Academy of Sciences, al. Lotnik\'{o}w 32/46, 02-668 Warszawa, Poland}
\author{M.~Sawicki}
\affiliation{Institute of Physics, Polish Academy of Sciences, al. Lotnik\'{o}w 32/46, 02-668 Warszawa, Poland}
\author{R. Jakie{\l}a}
\affiliation{Institute of Physics, Polish Academy of Sciences, al. Lotnik\'{o}w 32/46, 02-668 Warszawa, Poland}
\author{T. Dietl}
\affiliation{Institute of Physics, Polish Academy of Sciences, al. Lotnik\'{o}w 32/46, 02-668 Warszawa, Poland}
\affiliation{Institute of Theoretical Physics, Faculty of Physics, University of Warsaw, ul. Ho\.za 69, PL-00-681 Warszawa, Poland}
\affiliation{WPI-Advanced Institute for Materials Research (WPI-AIMR), Tohoku University, 2-1-1 Katahira, Aoba-ku, Sendai 980-8577, Japan}
\author{A. Navarro-Quezada}
\affiliation{Institut f\"ur Halbleiter- und Festk\"orperphysik, Johannes Kepler University, Altenbergerstr. 69, A-4040 Linz, Austria}
\author{B. Faina}
\affiliation{Institut f\"ur Halbleiter- und Festk\"orperphysik, Johannes Kepler University, Altenbergerstr. 69, A-4040 Linz, Austria}
\author{T. Li}
\affiliation{Institut f\"ur Halbleiter- und Festk\"orperphysik, Johannes Kepler University, Altenbergerstr. 69, A-4040 Linz, Austria}
\author{A. Bonanni}
\affiliation{Institut f\"ur Halbleiter- und Festk\"orperphysik, Johannes Kepler University, Altenbergerstr. 69, A-4040 Linz, Austria}
\author{J. Suf\mbox{}fczy\'{n}ski}
\email{Jan.Suffczynski@fuw.edu.pl}
\affiliation{Institute of Experimental Physics, Faculty of Physics, University of Warsaw, ul. Ho\.za 69, PL-00-681 Warszawa, Poland}



\begin{abstract}
 The question of the correlation between magnetization, band splittings, and magnetic circular dichroism (MCD) in the fundamental gap region of dilute magnetic semiconductors is examined experimentally and through model calculations, taking the case of wurtzite Ga$_{1-x}$Fe$_x$N as an example. Magnetization and polarization-resolved reflectivity measurements have been performed down to 2~K and up to 7~T for $x = 0.2$\%.  Optical transitions originating from all three free excitons $A$, $B$ and $C$, specific to the wurtzite structure, have been observed and their evolution with the magnetic field determined. It is demonstrated that the magnitude of the exciton splittings evaluated from reflectivity-MCD data can be overestimated by more than a factor of 2, as compared to the values obtained by describing the polarization-resolved reflectivity spectra with appropriate dielectric functions. A series of model calculations shows that the quantitative inaccuracy of MCD originates from a substantial influence of the magnetization-dependent exchange interactions not only on the spin splittings of excitons but also upon their linewidth and oscillator strength. At the same time, a method is proposed that allows to evaluate the field and temperature dependencies of the magnetization from MCD spectra. The accurate values of the excitonic splittings and of the magnetization reported here substantiate the magnitudes of the apparent $sp-d$ exchange integrals in (Ga,Fe)N previously determined.
\end{abstract}


%
\maketitle


\section{Introduction}
Excitonic magnetooptical phenomena in the fundamental gap region of dilute magnetic semiconductors (DMSs) are a prime source of information on the sp-d exchange splitting of electronic states\cite{Furdyna:1988_B, Ando:2000_B, Cibert:2008a_B, Gaj:2010_B} and spin dynamics,\cite{Crooker:2010_B} also in reduced dimensionality systems, such as nanowires\cite{Wojnar:2012_NL} and quantum dots.\cite{Schimpf:2012_JPChL,Cibert:2008_B} One of the widely studied magneto-optical effects is the magnetic circular dichroism (MCD) arising from the difference between optical absorption or reflectivity for left and right circular polarized light. The MCD intensity is defined as
\begin{equation}\label{MCD}
    MCD=\frac{I^+ -I^-}{I^+ +I^-},
\end{equation}
where $I^+$ and $I^-$ denote the $\sigma^+$ and $\sigma^-$ polarized components of the signal. In many DMSs, the MCD depends linearly on the exchange-induced splitting $\Delta$E of the free exciton states,\cite{Ando:2000_B} which, in turn, is proportional to the magnetization M of the localized spins.\cite{Gaj:1979_SSC} In the case of a single exciton line, quantitative information on $\Delta$E and, thus, on M can be obtained from the MCD intensity within the rigid shift approximation,\cite{Mason:2007_B} employed to evaluate the magnitudes of the sp-d exchange energies in various DMSs and their nanostructures.\cite{Kuno:1998_JChP, Hoffman:2000_SSC, Norris:2001_NL, Norberg:2006_JACS, Beaulac:2008_NL, Bussian:2009_NM}

In the case of wurtzite DMSs like (Ga,Fe)N or (Zn,Co)O, the description of the magnetooptical phenomena associated with free excitons is, however, more intricate.\cite{Ando:2001_JAP,Pacuski:2010_B, Schw03} Here, due to the effect of the trigonal crystal field and anisotropic spin-orbit coupling, the top of the valence band splits into three subbands.\cite{Dingle:1971_PRB} As a result, three free exciton transitions emerge in the spectral region near the band gap, and they are labeled as $A$, $B$ and $C$. They are close in energy and often not well resolved in the optical spectrum. Moreover, at a given circular polarization, the magnetization-induced shift of exciton $B$ is opposite to the ones of excitons $A$ and $C$.\cite{Pacuski:2006_PRB,Pacuski:2008_PRL, Pacuski:2011_PRB,Suffczynski:2010_PRB} It has been suggested\cite{Ando:2003_APL} that this may even lead to the mutual cancellation of the magnetooptical effects associated with the excitons $A$ and $B$. Furthermore, the electron-hole exchange interaction leads to a mixing between $A$ and $B$ excitonic states, resulting in an anticrossing that affects the magnetization-induced shifts of the excitonic states.\cite{Pacuski:2006_PRB} These factors can make that the relation between the magnitudes of the MCD, exciton splitting, and magnetization is not longer linear in wurtzite DMSs.

In this work we report on magneto-reflectivity and magnetization measurements carried out on paramagnetic (Ga,Fe)N layers, whose high crystalline quality has been demonstrated by a range of structural characterization methods.\cite{Bonanni:2007_PRB,Bonanni:2008_PRL} We aim at clarifying,  in the case of wurtzite DMSs with low carrier densities and randomly distributed magnetic ions, the relation between the magnitudes of (i) free exciton splittings, (ii) MCD obtained from magneto-reflectivity, and (iii) magnetization. From near band gap reflectivity data we determine the MCD spectra as well as the exciton splittings by fitting the magneto-reflectivity spectra with an appropriate dielectric function. An excellent proportionality is found between the field dependence of the  magnetization and the shift of the exciton $A$ line in $\sigma^-$ polarization. In contrast, the degree of proportionality between the magnitudes of the MCD and of the magnetization depends on the choice of the method for the determination of the MCD magnitude from the MCD spectra. We indicate a method assuring the highest degree of proportionality. Furthermore, we show that despite the opposite signs of the exchange splittings, the contributions of the excitons $A$ and $B$ to the MCD do not cancel out. We explain this fact presenting a series of MCD simulations demonstrating that the MCD signal depends not only on the magnitude of the exchange-induced exciton splittings but also on changes of the exciton linewidth and oscillator strength with the magnetic field. It is expected that our conclusions apply to a wide class of wurtzite DMSs, {\em e.~g.}, nitrides and oxides like (Ga,Mn)N and (Zn,Co)O. Our results substantiate also the magnitudes of the apparent $sp-d$ exchange integrals determined previously for (Ga,Fe)N.\cite{Pacuski:2008_PRL}

It is worth noting that the breakdown of a simple relation between the magnitudes of the MCD, $\Delta$E, and M has also been found in other situations, for instance in the temperature studies of the MCD at photon energies E near the fundamental absorption edge in antiferromagnetic MnTe (Ref.~\onlinecite{Ando:1992_PRB}) and ferromagnetic (Ga,Mn)As (Refs. ~\onlinecite{Beschoten:1999_PRL, Dietl:2001_PRB}). It was pointed out\cite{Ando:1992_PRB, Dietl:2001_PRB} that a proper interpretation of the MCD should actually take into account the spectral dependence of the absorption coefficient $\alpha$, leading, in the linear approximation in $\Delta$E, to the MCD of the form,\cite{Cibert:2008a_B, Ando:1992_PRB}
\begin{equation}\label{AndoMCD}
MCD=-\frac{\Delta E}{2} \frac{1}{\alpha} \frac{d\alpha}{dE}.
\end{equation}
This approximation can be useful for describing MCD associated with a splitting of a single line or a single edge but it breaks down in more complex situations, including the case of overlapping exciton lines.

\section{Samples and experiment}
The studied $0.24~\mu$m thick (Ga,Fe)N layers are grown by metal-organic vapor phase epitaxy (MOVPE) on a $1~\mu$m thick GaN buffer deposited on a [0001] sapphire substrate, which results in single crystal layers with the $c$-axis perpendicular to the film plane.\cite{Bonanni:2007_PRB} According to previous electron paramagnetic resonance (EPR) studies,\cite{Bonanni:2007_PRB} and as confirmed by the observation of photoluminescence (not shown) corresponding to intracenter $^{4}$T$_{1}$(G) $\rightarrow$ $^{6}$A$_{1}$(S) transitions at 1.3 eV,\cite{Malguth:2006_PRBb} the iron in the studied samples is present predominantly as Ga-substitutional Fe$^{3+}$ ions in the high spin $S = 5/2$ configuration. The Fe concentration resulting from magnetometry measurements, estimated as a difference between the magnetization values~\cite{Pacuski:2008_PRL} measured at 1.8 and 5 K, is $x = 0.2$~\%.

Reflectivity measurements are performed at pumped helium temperatures ($T \simeq 2$~K), in a magnetic field $B$ up to $7$~T applied in the Faraday configuration (magnetic field parallel to the direction of the light propagation) and with the light propagation along the $c$ axis of the film. The sample is illuminated by unpolarized white light at normal incidence. The signal is detected by a Peltier cooled CCD camera coupled to a grating monochromator ($2400$ grooves/mm). A set of a quarter wave plate and a linear polarizer inserted in the signal path enables the detection of the signal at both circular polarizations $\sigma^+$ and $\sigma^-$.

Magnetization measurements are carried out in a superconducting quantum interference device (SQUID) as a function of a magnetic field up to 7~T applied perpendicularly and in parallel with respect to the sample growth axis at temperatures from 2 to 300~K, according to the procedure outlined recently.\cite{Sawicki:2011_SST}

\begin{figure}
   \includegraphics[width=0.9\linewidth]{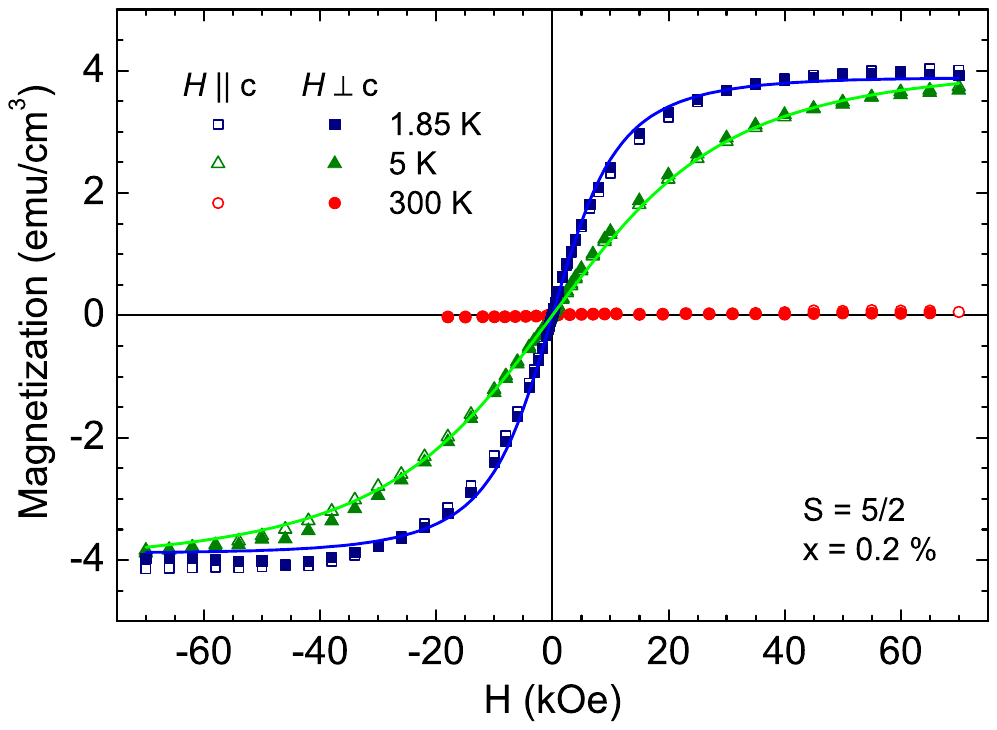}\\
  \caption{(color online) Magnetization in the magnetic field up to 7 T applied parallel (open points) or perpendicularly (full points) to the growth axis at at 2, 15, and 300~K for a sample of Ga$_{1-x}$Fe$_x$N. Lines are dependencies calculated following Eq.~\ref{Eq:magnetization} for non-interacting ions with $S = 5/2$ and concentration $x = 0.2$~\%.}\label{magnetometry}
\end{figure}

High-resolution transmission electron microscopy (HRTEM) measurements performed on cross-sectional specimens of the samples prepared by a standard mechanical polishing followed by Ar+ ion milling at 4\ kV, did not reveal iron-rich precipitations in the studied films, consistent with the absence of a ferromagnetic component in the magnetization data.

\section{Results of magnetization measurements}
\label{sec:magnetometry}
In Fig.~\ref{magnetometry} the magnetization in the magnetic field perpendicular and parallel to the sample growth axis is reported. No magnetic anisotropy is observed and the data are well described by a paramagnetic isotropic Brillouin function for $S = 5/2$,
\begin{equation}\label{Eq:magnetization}
M(T,H) = g\mu_{\text{B}}xN_0S\text{B}_S(T,H),
\end{equation}
confirming that the uncoupled Fe$^{3+}$ ions determine the magnetic properties in the studied samples. Here, $g=2.0$, $N_0$ is the cation concentration, and the Fe concentration is $x = 0.2$~\% for the data presented in Fig.~\ref{magnetometry}.

\section{Reflectivity spectra and the determination of exchange integrals}
\label{sec:reflectivity}
\begin{figure}
    \includegraphics[width=0.8\linewidth]{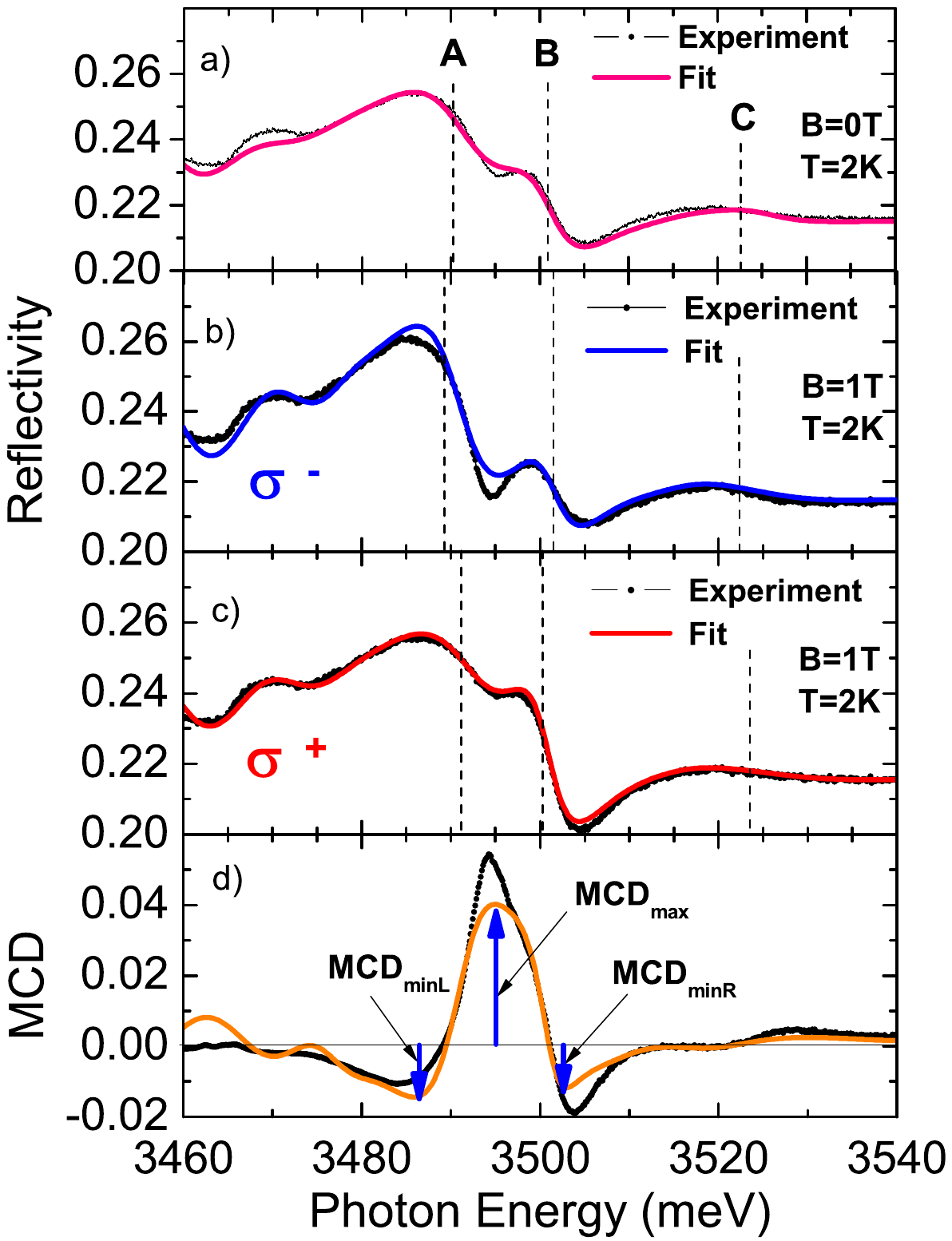}
  \caption{(color online) Experimental (points) and fitted (solid lines) reflectivity spectra at $ B = 0$~T (a) and at 1~T for $\sigma^-$ (b) and $\sigma^+$ (c) polarizations. The energies of optical transitions corresponding to the excitons $A$, $B$ and $C$ are indicated (vertical dashed lines). In panel (d) the experimental (points) and modeled (lines) MCD spectra are given, as determined from the reflectivity spectra shown in panels (b) and (c).}\label{widma1T}
\end{figure}

In Fig.~\ref{widma1T}(a) a typical reflectivity spectrum of the studied samples in the near bandgap spectral region at zero magnetic field is presented. The presence of all three $A$, $B$ and $C$ excitonic transitions, as indicated in the plot, confirms the high crystal quality of the studied layers.  A giant polarization-dependent shift of the excitonic transitions is observed in the magnetic field, as shown in Figs.~\ref{widma1T}(b) and \ref{widma1T}(c). The $A$ and $B$ excitons shift in opposite directions and are spectrally well resolved at $\sigma^-$ polarization (Fig.~\ref{widma1T}(b)), while they gradually merge with the increase of the magnetic field in $\sigma^+$ polarization (Fig.~\ref{widma1T}(c)).

Since the magnitudes of the excitonic shifts and linewidths are comparable, in order to determine accurately the magnitude of the exciton splittings in the magnetic field we describe the entire experimental reflectivity spectra by appropriate response functions. The model takes into account the refractive index of each layer and the multiple reflections of the light in the three layer structure. First, the reflectivity coefficients are calculated for each interface between adjacent layers as,

\begin{equation}\label{refcoefter}
    r^{\pm}_{i,i+1}=\frac{\sqrt{\epsilon^{\pm}_{i}}-\sqrt{\epsilon^{\pm}_{i+1}}}{\sqrt{\epsilon^{\pm}_{i}}+\sqrt{\epsilon^{\pm}_{i+1}}},
\end{equation}
where $\epsilon^{\pm}_i$ is the energy dependent, complex dielectric function of the $i$-th layer in $\sigma^+$ and $\sigma^-$ polarization, respectively. The $\epsilon^{\pm}_1$ and $\epsilon^{\pm}_2$ are determined for the (Ga,Fe)N layer and GaN buffer respectively taking into account the $A$, $B$ and $C$ free exciton transitions, the excited states of the excitons, and the continuum of unbound states within each layer. The energy positions, linewidths and amplitudes of the $A$, $B$ and $C$ excitonic transitions are treated as adjustable parameters, while the values of the remaining material parameters are taken from Ref.~\onlinecite{Pacuski:2008_PRL}. The dielectric constant of the sapphire substrate  $\epsilon^{\pm}_3 = 3.29$ is assumed to be frequency independent. Then, the coefficient of reflectivity of the whole structure is calculated for the two circular polarizations of the light $\sigma^+$ and $\sigma^-$ as $R^{\pm} = |r^{\pm}|^2$, where
\begin{equation}\label{refcoef}
     r^{\pm}=\frac{r^{\pm}_{01}+r^{\pm}_{123}e^{2i\beta_1^{\pm}}}{1+r^{\pm}_{01}r^{\pm}_{123}e^{2i\beta_1^{\pm}}},
\end{equation}
\begin{equation}\label{refcoefbis}
    r^{\pm}_{123}=\frac{r^{\pm}_{12}+r^{\pm}_{23}e^{2i\beta_2^{\pm}}}{1+r^{\pm}_{12}+r^{\pm}_{23}e^{2i\beta_2^{\pm}}}
\end{equation}
and $\beta^{\pm}_i=(\omega/c)l_i \sqrt{\epsilon^{\pm}_i}$ is the dephasing of the electromagnetic wave of frequency $\omega$ passing through the $i-th$ layer of thickness $l_i$. Comparing to the previous model of magneto-reflectivity in (Ga,Fe)N (Ref.~\onlinecite{Pacuski:2008_PRL}), we take into account Fabry-P\'{e}rot interferences occurring in our structure consisting of thin layers but neglect polaritonic effects\cite{Hopfield:1963_PR} that would be more important in the case of spectrally narrower transitions and smaller Zeeman splittings than in our case.

\begin{figure}
   \includegraphics[width=0.8\linewidth]{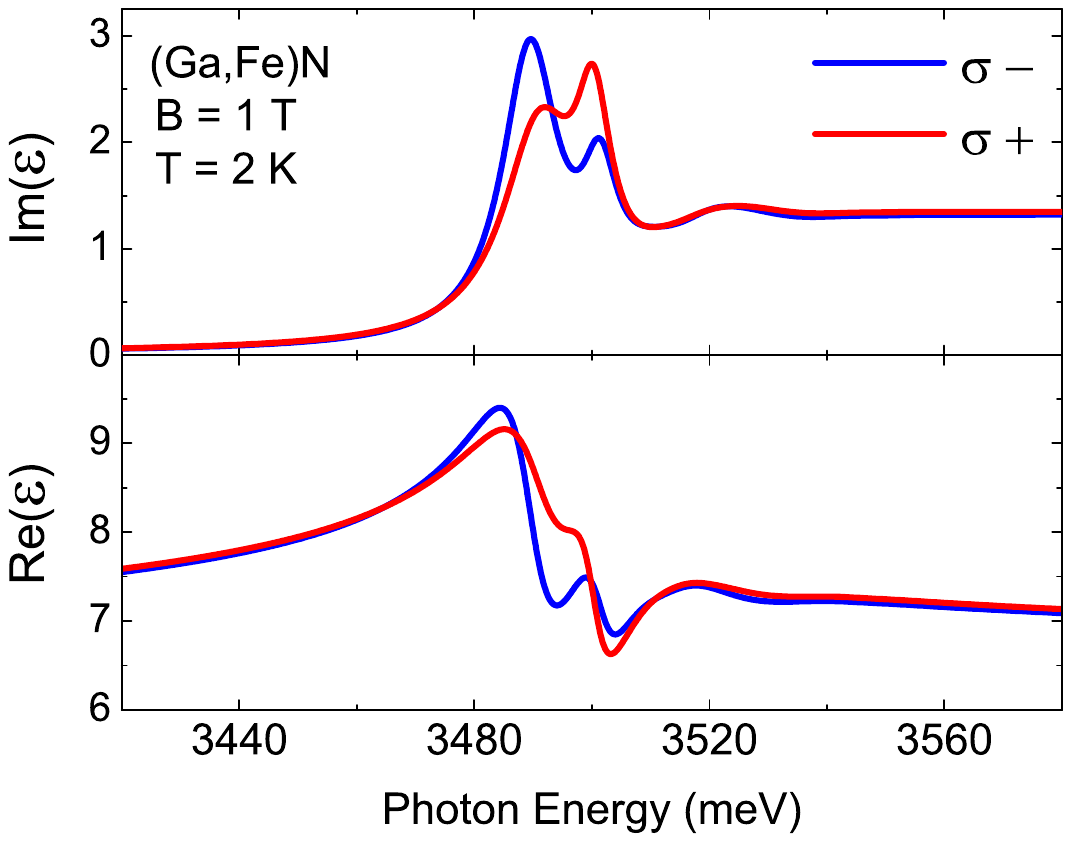}\\
  \caption{(color online) The calculated real and imaginary parts of the (Ga,Fe)N ($x = 0.2$~\%) dielectric function in B = 1 T and at T = 2 K for both, $\sigma^+$ and $\sigma^-$, circular polarizations of the light.}\label{dielfunc}
\end{figure}

\begin{figure}
  \includegraphics[width=1\linewidth]{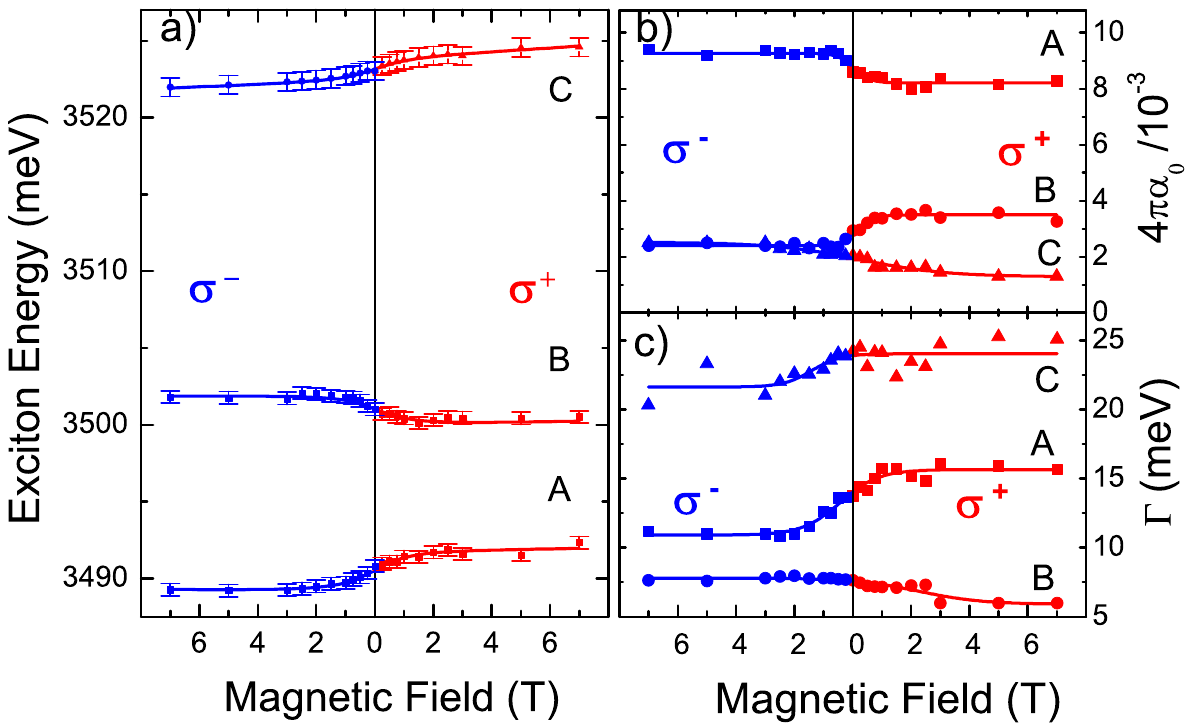}
  \caption{(color online) Energy positions (a), linewidths (b), and  the oscillator strength (c) of the $A$, $B$ and $C$ excitons for both circular polarizations (points), as obtained by fitting the reflectivity spectra shown in Fig.~\ref{widma1T}. The solid lines in (b) and (c) are guides to the eye(a), whereas in (a) they represent the model fit with two fitting parameters, the \emph{apparent} $s-d$ and $p-d$ exchange energies for the conduction and valence band, respectively.}
  \label{fitexc}
\end{figure}

The computed real and imaginary parts of the (Ga,Fe)N dielectric function in 1 T and at 2 K are plotted in Fig.~\ref{dielfunc}. The corresponding reflectivity spectra are shown together with the experimental results in Figs.~\ref{widma1T}(b) and \ref{widma1T}(c). A  good match of the experimental data and modeling is evident in the whole near-band-gap spectral region. The energy positions, linewidths, and oscillator strengths of the $A$, $B$ and $C$ excitonic transitions as a function of the magnetic field, as obtained from the fitting procedure, are presented in Fig.~\ref{fitexc}.

The obtained dependencies of the exciton energies on the magnetic field, shown in Fig.~\ref{fitexc}(a), provide information on the $s-d$ and $p-d$ exchange splitting of the conduction and valence bands in (Ga,Fe)N at a given magnetization of the Fe ions, $M(T,H)$. We interpret these dependencies by taking into account the electron-hole exchange interaction and by incorporating into the $kp$ Hamiltonian for the wurtzite structure the $sp-d$ exchange terms, $I\bm{sM}(T,H)/g\mu_{\text{B}}$, where the exchange integrals are $I = \alpha^{\text{(app)}}$ and $I =\beta^{\text{(app)}}$ for the $s$-type conduction band and for the $p$-type valence band, respectively.\cite{Pacuski:2008_PRL} The index "app" underlines that the exchange integrals introduced in this way may be renormalized by corrections to the virtual and molecular field approximations.\cite{Dietl:2008_PRB} The apparent exchange energies we determine from our magneto-reflectivity data (Fig.~\ref{fitexc}) are $N_0 \alpha^{\text{(app)}} = -0.05 \pm 0.1$~eV and $N_0 \beta^{\text{(app)}} = +0.5 \pm 0.2$~eV, respectively. The main contribution to the error is related to the uncertainty of the iron concentration determination.

The value of $N_0 \beta^{\text{(app)}}$ confirms its previous determination made in  Ref.~\onlinecite{Pacuski:2008_PRL} ($N_0 \alpha^{\text{(app)}} = +0.1 \pm 0.2$~eV and $N_0 \beta^{\text{(app)}} = +0.5 \pm 0.2$~eV) and is consistent with the results of recent optical works on other wide band-gap DMS, like (Ga,Mn)N (Refs.~\onlinecite{Pacuski:2007_PRB,Suffczynski:2010_PRB}) and (Zn,Mn)O.\cite{Przezdziecka:2006_SSC,Pacuski:2011_PRB} Remarkably, the values of $N_0 \beta^{\text{(app)}}$ found by exciton spectroscopy for nitrides and oxides have an opposite sign and smaller magnitudes than expected from photoemission studies\cite{Hwang:2005_PRB} and from chemical trends implied by previous works on magnetically doped chalcogenides.\cite{Furdyna:1988_B,Gaj:2010_B}  These  puzzling findings were explained\cite{Dietl:2008_PRB} by considering the influence of the antiferromagnetic $p-d$ exchange on the band states in a nonperturbative way. The strong coupling effects are particularly relevant in the case of nitrides and oxides, where -- owing to the short bond length -- the $p-d$ hybridization is large. This approach demonstrates that if the potential brought about by the transition metal impurity is deep enough to bind a hole with an antiparallel spin, the corresponding extended states are pushed out of the magnetic impurity, as they must be orthogonal to the bound state. In particular, the theory anticipates a sign reversal of the $p-d$ exchange integral describing the giant spin splitting of the valence band states, as observed for (Ga,Fe)N.
It is interesting to note that also the magnitude of the $s-d$ exchange energy is reduced in comparison to the II-VI DMSs. This might indicate that the poking of the exciton hole out of the magnetic ion reduces also the overlap of the exciton electron with localized spins.

\section{MCD {\em vs.} magnetization}
\label{sec:MCD}
\begin{figure}
  \includegraphics[width=0.9\linewidth]{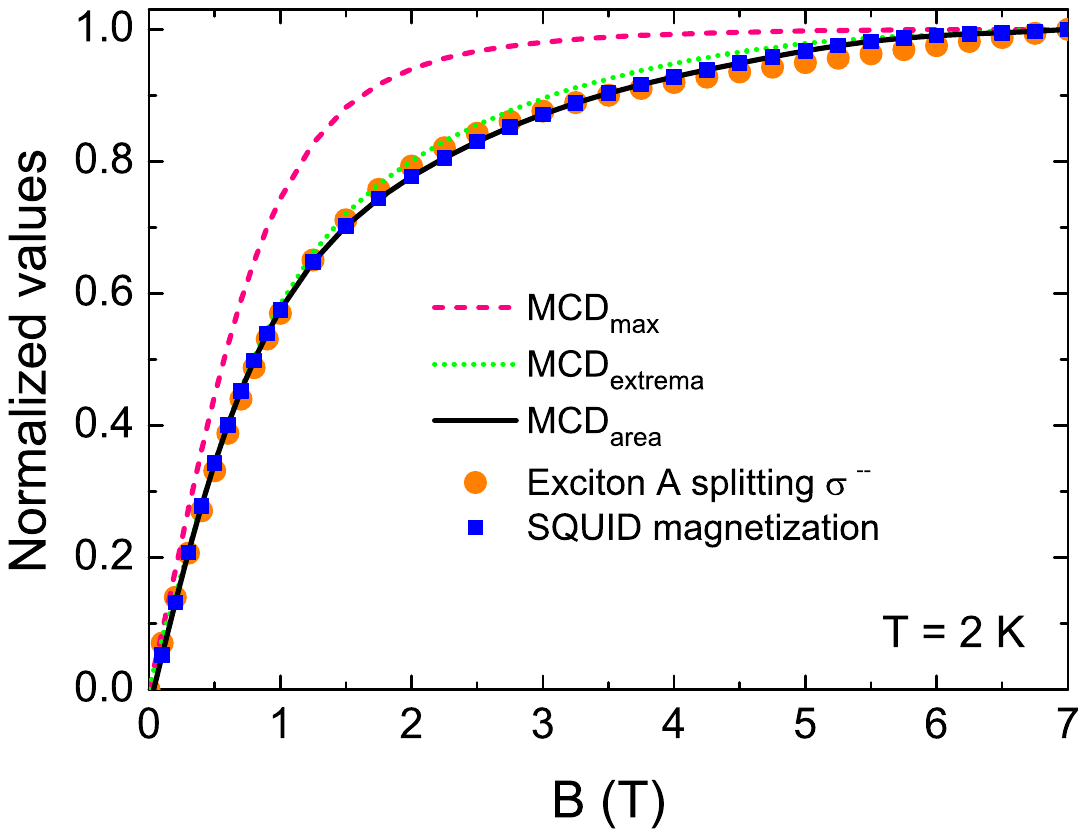}
  \caption{(color online) Amplitude of the main MCD peak MCD$_{\textbf{max}}$ (dashed line), the sum of the absolute values of the MCD extrema MCD$_{\textbf{extrema}}$ (dotted line), integrated area under the MCD curve MCD$_{\textbf{area}}$ (solid line) plotted together with the exciton $A$ shift in polarization $\sigma^-$ (circles) and magnetization (squares), all normalized by their values at 7~T.}
  \label{splitinteB}
\end{figure}

In zinc-blende DMSs with degenerate valence bands at $k=0$ and, thus, with a single free exciton transition, the splitting is proportional to the magnetization.\cite{Furdyna:1988_B} In wurtzite DMSs, three spectrally close excitonic transitions contribute to the reflectivity spectra near the band gap. In the case of (Ga,Fe)N, only exciton $A$ in $\sigma^-$ polarization, whose energy decreases with the magnetic field, does not interact with other states.\cite{Pacuski:2008_PRL} As shown in Fig.~\ref{splitinteB}, the redshift of exciton $A$ in $\sigma^-$ polarization and the magnetization show the same dependence on the magnetic field. In general, however, due to the anticrossings between excitonic states occurring when their exchange-induced shifts increase (see Fig.~\ref{fitexc}), the excitons exchange splittings cease to be proportional to the magnetization. A question then arises on the nature of the relation between MCD and magnetization in such a case.

The experimental MCD and modeled MCD (MCD$_{\textbf{fit}}=\frac{R^+ -R^-}{R^+ +R^-}$) are shown for (Ga,Fe)N at B = 1~T as a function of the photon energy in Fig.~\ref{widma1T}(d). As seen, the MCD spectrum exhibits a wide pronounced maximum (MCD$_{\textbf{max}}$) in the spectral region of excitons $A$ and $B$, centered around 3495~meV. The MCD$_{\textbf{max}}$ is accompanied by two minima (MCD$_{\textbf{minL}}$ and MCD$_{\textbf{minR}}$) on its sides, as marked in Fig.~\ref{widma1T}(d). The MCD signal is  weak outside this region, indicating that the $A$ and $B$ excitons provide the main contribution to the magnitude of near band-gap MCD in (Ga,Fe)N.

The shape of the MCD spectrum shown in Fig.~\ref{widma1T}(d), although resulting from the contributions of three excitons, is actually similar to the typical spectra of zinc-blende DMSs.\cite{Ando:2000_B,Ando:2004_JPCM} It is interesting to analyse which characteristics of the MCD spectrum in (Ga,Fe)N would assure the highest degree of agreement with magnetization. We consider three quantities: 1) the integrated absolute value of the MCD intensity\cite{Hochst:2002_PRB} in the excitonic region, MCD$_{\textbf{area}}$; 2) the maximum of the MCD intensity, MCD$_{\textbf{max}}$; 3) the sum of the absolute values of the MCD extrema, MCD$_{\textbf{extrema}}$ = |MCD$_{\textbf{minL}}$| + MCD$_{\textbf{max}}$ + |MCD$_{\textbf{minR}}$|.

\begin{figure}
  \includegraphics[width=0.9\linewidth]{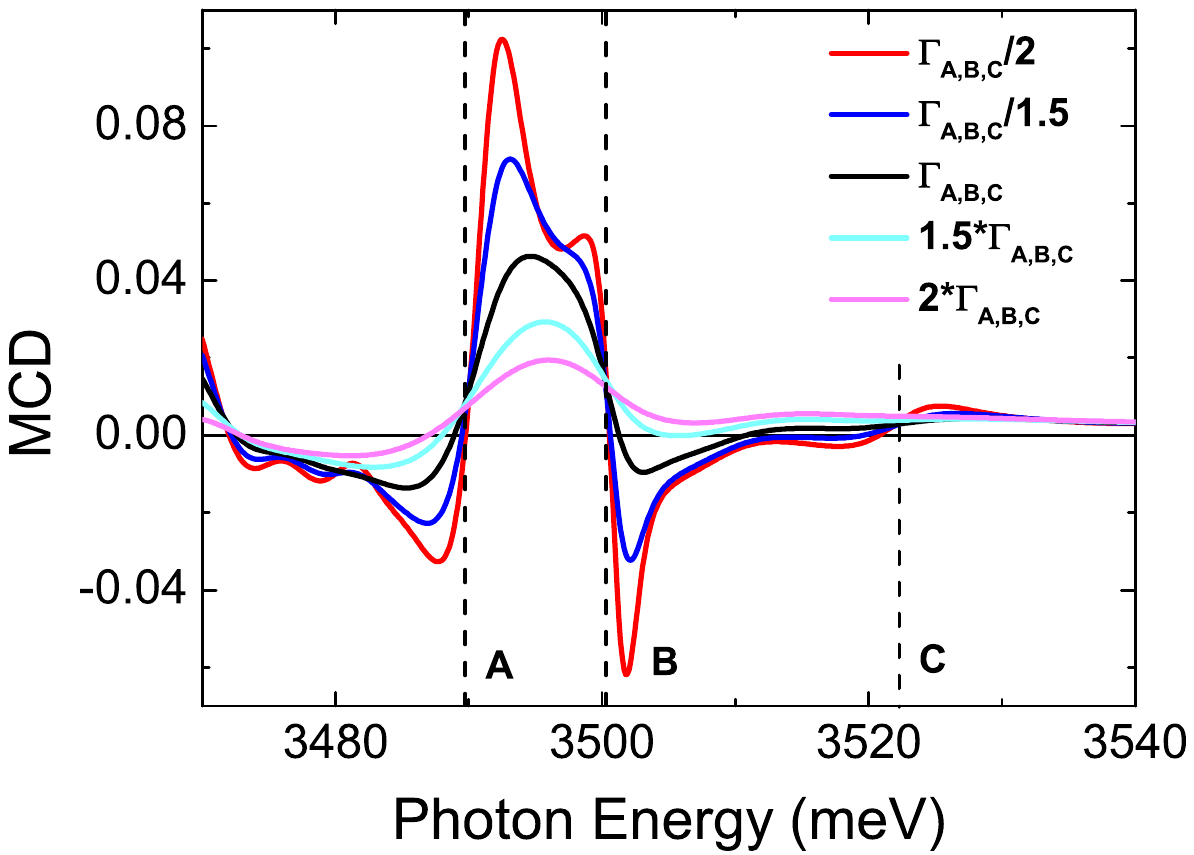}
  \caption{(color online) Calculated MCD spectra at 1~T for various linewidths of the excitons $A$, $B$ and $C$, changed be a factor between 1/2 and 2. The vertical lines indicate the positions of the excitons in the absence of a magnetic field.}
  \label{wavesMCDvsGamma}
\end{figure}

The side extrema (|MCD$_{\textbf{minL}}$| and |MCD$_{\textbf{minR}}$|) and the central maximum (MCD$_{\textbf{max}}$) correspond respectively to the minima and the sum of maxima of the lorentzian type contributions to the MCD curve from the $A$ and $B$ excitons (as seen in Figs.~\ref{widma1T} and ~\ref{wavesMCDvsGamma}). Thus, the sum of the extrema of the MCD values should provide reasonable information on the MCD magnitude. A similar method (summation of curve extrema) was applied for the determination of  the magnitude of magneto-optic Kerr effect in (Cd,Mn)Te.~\cite{Testelin:1997_PRB}

The three considered quantities normalized to their values at 7 T are plotted, together with the magnetization, in Fig.~\ref{splitinteB}. As seen, all quantities as a function of the magnetic field exhibit a Brillouin-like dependence and undergo saturation. An excellent agreement is observed between the values of MCD$_{\textbf{area}}$ and the magnetization. This is not the case, however, of the MCD$_{\textbf{max}}$, which deviates significantly from the magnetization. We have checked that for the studied layers, in contrast to zinc-blende DMSs,\cite{Norberg:2006_JACS,Ando:2000_B} there is no specific photon energy for which the MCD intensity is proportional to the magnetization. The agreement between the MCD magnitude and the magnetization can be, however, restored after adding to MCD$_{\textbf{max}}$ the magnitudes of the side extrema |MCD$_{\textbf{minL}}$| and |MCD$_{\textbf{minR}}$|.

It can be expected that the magnetooptical methods of magnetization determination will break down when the exciton linewidth becomes larger than either the splitting of the excitonic states or of the energy difference between particular excitons. In order to determine the range of excitonic linewidths assuring an agreement between MCD and magnetization, we calculate the MCD spectra varying the linewiths of the excitons $A$, $B$ and $C$ by a factor from 1/2 to 2 with respect to their original value at 1~T. When the excitonic linewidths increase, the contributions of the excitons $A$ and $B$ are no longer resolved in the reflectivity spectrum. At the same time, as seen in Fig.~\ref{wavesMCDvsGamma}, a less intense and broadened MCD spectrum is expected. It is clearly visible, however, that the MCD signals originating from excitons $A$ and $B$ do not cancel out, as suggested previously,\cite{Ando:2001_JAP} but rather add. Actually, despite the opposite splittings, a difference in the energy of excitons $A$ and $B$ makes that they contribute with the same sign to the MCD signal.

\begin{figure}
  \includegraphics[width=0.9\linewidth]{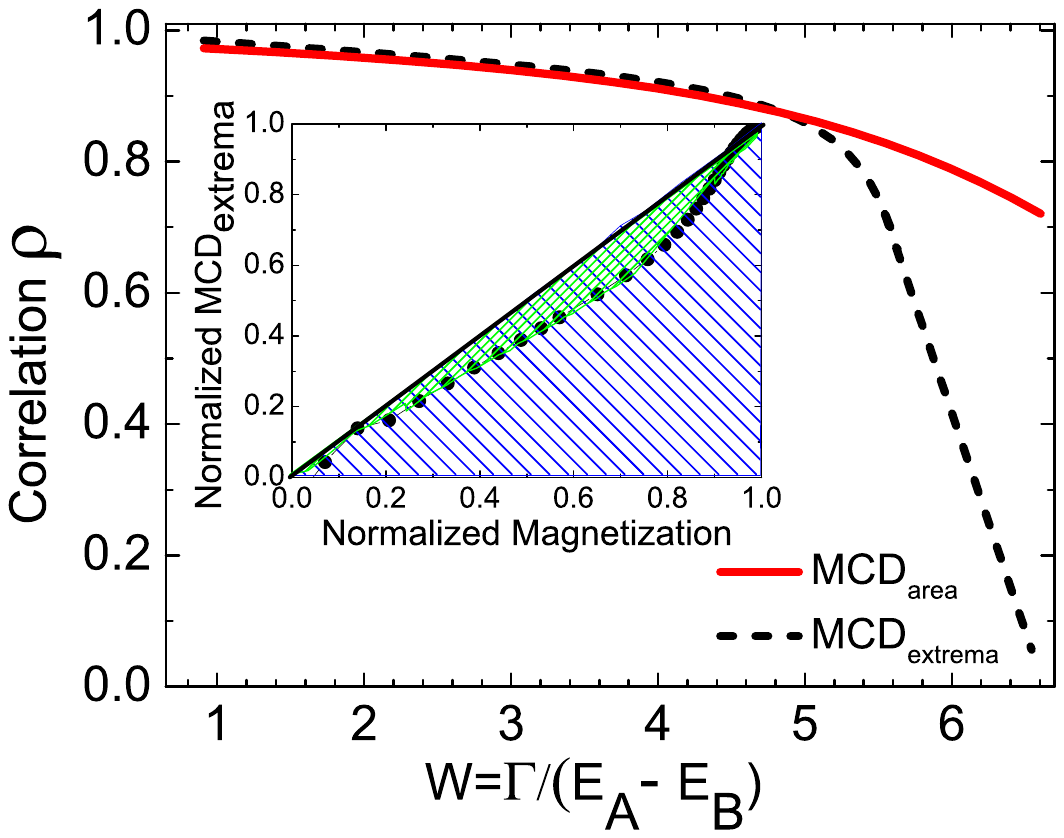}
  \caption{(color online) Coefficient $\rho$ (Eq.~\ref{def:rho}) describing the correlation between the magnitudes of magnetization and MCD for MCD$_{\textbf{area}}$ (solid line) and MCD$_{\textbf{extrema}}$ (dashed line), as defined in Sec.~\ref{sec:MCD}, plotted vs. the parameter $W$ (Eq.~\ref{def:W}) describing the spectral broadening of the excitonic transitions. Good agreement between the magnitude of magnetization and MCD ($\rho$ > 0.9) is maintained up to $W\sim 5$ and deteriorates for higher $W$, particularly in the case of MCD$_{\textbf{extrema}}$. Inset: normalized MCD$_{\textbf{extrema}}$ vs. normalized magnetization for $W = 3.8$. The correlation coefficient $\rho$ is calculated as the ratio of the triangle areas below the dotted and solid lines.}
  \label{correl_fct_ratio}
\end{figure}

In order to quantify the degree of agreement between MCD and magnetization, we define a correlation coefficient $\rho$ as
\begin{equation}\label{def:rho}
\rho=1-\frac{\int_0^1\sqrt{(y-z)^2} \,dz}{\int_0^1 z \,dz},
\end{equation}
where $y$ and $z$ are the  magnitudes of MCD and magnetization, normalized by their values at 7~T. The coefficient $\rho$ corresponds to the ratio of the area below the dotted curve to the area below the dashed line shown in the inset to Fig.~\ref{correl_fct_ratio}, and attains a value of 1 for a perfect agreement between normalized MCD and magnetization.%

The values of $\rho$ are plotted in Fig.~\ref{correl_fct_ratio} as a function of the relative broadening $W$:

\begin{equation}\label{def:W}
W=\frac{\Gamma_A}{(E_{B} - E_{A})},
\end{equation}

where $\Gamma_i$ and $E_i$ are the linewidth and energy of the $i$-th exciton in the absence of a magnetic field. According to the results displayed in Fig.~\ref{correl_fct_ratio}, for  $W\approx 5$ (the case of narrow exciton lines) the magnitudes of integrated MCD and the sum of the MCD extrema agree reasonably well ($\rho > 0.9$) with the magnetization. However, with increasing excitonic broadening $\Gamma_A$ the degree of correlation $\rho$ diminishes. For $W>5$, $\rho$ calculated for the sum of the MCD extrema drops rapidly, showing that MCD$_{\textbf{extrema}}$ is no longer proportional to the magnetization. The influence of the exciton broadening on $\rho$ is less pronounced in the case of MCD$_{\textbf{area}}$. Since in a typical case $E_{B} - E_{A} \approx 10$~meV, even if $\Gamma_A$ is as large as 50 meV, MCD$_{\textbf{area}}$ still describes correctly the normalized magnitude of the magnetization.

\section{Effects of exciton splitting, linewidth, and oscillator strength on MCD}
It is well established\cite{Maslana:2001_PRB, Pacuski:2006_PRB} that the $sp-d$ exchange interaction affects not only the energy of exciton transitions but also their linewidth and oscillator strength. It was shown that in (Cd,Mn)Te a greater contribution to the Faraday rotation may come from the polarization-dependent oscillator strength than from the energy shift.\cite{Maslana:2001_PRB}

\begin{figure}
  \includegraphics[width=0.9\linewidth]{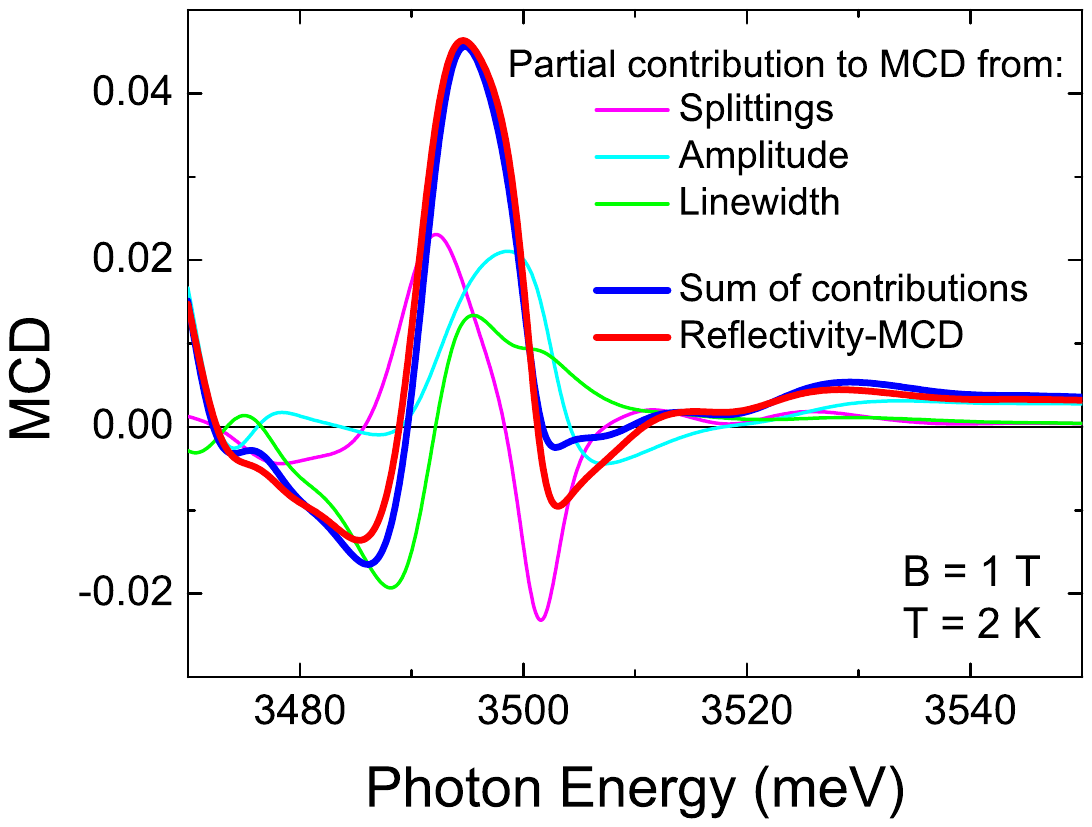}
  \caption{(color online) Contributions to the MCD spectrum at 1~T, originating from the exchange-induced variation of excitonic shift, linewidth, and oscillator strength calculated by varying a given type of parameter (e.g., splitting) of the $A$, $B$ and $C$ excitons and freezing the remaining two at their zero-field values. The sum of the partial contributions is seen to be in agreement with the magnitude of the MCD as determined from the reflectivity spectra, MCD$_{\textbf{fit}}$.}
  \label{contrib1T}
\end{figure}

The dependence of the exciton shift, linewidth, and oscillator strength on the sense of circular polarization (Fig.~\ref{fitexc}) determined here, allows us to establish which of these parameters gives a dominant contribution to MCD in (Ga,Fe)N. To do so we calculate the contributions to MCD coming from the excitonic shift, linewidth, and oscillator strength by letting only one parameter ({\em e.g.}, the shift of the $A$, $B$ and $C$ excitons) to vary with the magnetic field while freezing the remaining two at their zero-field values.  As shown in Fig.~\ref{contrib1T}, each of these three excitonic characteristics contributes sizably to the total MCD signal.

The simulations performed as a function of the magnetic field show that none of the three parameters characterizing the excitons gives a contribution to the MCD that would itself be proportional to the magnetization. This means that the observed MCD is a convolution of all three partial contributions and, in particular, that individual $A$, $B$ and $C$ excitonic splittings would not lead to an MCD signal proportional to the magnetization. This result, specific to wurtzite DMSs, is a direct consequence of the close spectral vicinity of the excitonic transitions,  non-zero linewidths, and opposite shifts of the excitons $A$ and $B$ in the magnetic field.

Finally, we comment on the reliability of the determination of the splitting from reflectivity-MCD. A comparison between the different contributions to the overall MCD (see Fig.~\ref{contrib1T}) indicates that the determination of the excitonic splitting when neglecting the contributions related to excitonic linewidth and oscillator strength would lead to the values overestimated by a factor of at least two. This means that in order to evaluate the excitonic splittings and thus the exchange energies meaningfully, a fitting of reflectivity or absorption spectra at various polarizations, as performed in the present work, is necessary.

\section{Conclusions}
More specifically, our work has provided:

(i) the experimental determination and modeling of MCD in (Ga,Fe)N, not reported previously;

(ii) the quantitative analysis of MCD in the case of three overlapping exciton lines, specific to the wurtzite structure;

(iii) the explicit evaluation of the contributions to MCD originating from the effect of a magnetic field and light polarization upon the exciton splitting,  linewidth, and oscillator strength;

(iv) the insight that the presence of various contributions to MCD precludes the determination of the excitonic splittings from the MCD magnitude in the case of wurtzite DMSs;

(v) the elucidation that there is no cancellation of the contributions to MCD originating from opposite splittings of excitons $A$ and $B$ in a magnetic field, in contrast to common sense predictions; (see Ref.~\onlinecite{Ando:2003_APL})

(vi) the demonstration that the integrated MCD amplitude describes the magnitude of (Ga,Fe)N magnetization much better than the MCD intensity at any particular wavelength. It is to be found out whether this observation is related to the recently proposed dichroic f-sum rule for magnetized insulators. (see Ref.~\onlinecite{Souza:2008_PRB})

The above conclusions are expected to be valid also in the case of other wurtzite DMSs with three excitons overlapping spectrally, the case of, e. g., magnetically doped nitrides and oxides.

\section*{Acknowledgments}
We acknowledge the support by the European Commission through the FunDMS Advanced Grant (No 227690) of the ERC within the "Ideas" 7th Framework Programme, by the Austrian Fonds zur F\"{o}rderung der wissenschaftlichen Forschung-FWF (P22477, P20065 and P24471), and by the Polish NCBiR project LIDER.


\end{document}